# Measuring the knowledge base of regional innovation systems in Germany in terms of a Triple Helix dynamics

*Research Policy* (forthcoming)

Loet Leydesdorff [1] & Michael Fritsch [2]

**Abstract**

The interaction among the three sub-dynamics of economic exchange, technological innovation, and institutional control can be captured with a generalized Triple Helix model. We propose to use the information contained in the configuration among the three sub-dynamics as an indicator of the synergy in a configuration. This measure indicates the reduction of the uncertainty which prevails at the level of an innovation system. On the basis of data at the district level in Germany, the conclusions of a previous study about the Netherlands are tested: medium-tech manufacturing is the main driver of the knowledge-based configuration in a regional economy, while knowledge-intensive services tend to uncouple the economy from the regional configuration. At the level of regions (NUTS-2) Germany's knowledge-based economy is no longer structured in terms of the previous East-West divide of the country, while this divide has prevailed at the level of the states (NUTS-1) that constitute the Federal Republic. However, the effects of high and medium-tech are not specific for the western or eastern parts of the country. The configuration of medium-tech manufacturing can be considered a better indicator of the knowledge-based economy than that of high-tech manufacturing.

---

[1] (corresponding author) Amsterdam School of Communications Research, University of Amsterdam, Kloveniersburgwal 48, 1012 CX Amsterdam, The Netherlands; loet@leydesdorff.net, http://www.leydesdorff.net ; fax: +31-20-5253681.

[2] Technical University Bergakademie Freiberg, German Institute for Economics (DIW), Berlin, and Max Planck Institute of Economics, Jena, Germany. Address: Technical University Bergakademie Freiberg, Faculty of Economics and Business Administration, Lessingstr. 45, D-09596 Freiberg, Germany; michael.fritsch@tu-freiberg.de

## 1. Introduction

One of the main ideas behind the concept of innovation *systems* is that innovation takes place both within firms and across the interfaces among institutional agents like universities, industries, and government agencies. Innovation systems differ in terms of how the fluxes through the networks are integrated and whether the heterogeneous fluxes (economic exchange relations, novelty production, and organizational control) provide opportunities for synergy. The networks provide only the knowledge infrastructure, while the knowledge base of an innovation system is shaped by a division of innovative labor at the national and/or regional level given such an infrastructure. The synergy between the industrial structure, geographical distributions, and academic traditions can be considered crucial for the strength of an innovation system (Fritsch, 2004).

The Triple Helix of university-industry-government relations has hitherto been developed mainly as a *neo-institutional* model for studying the network arrangements among these agents (Etzkowitz *et al*., 2000; Powell & DiMaggio, 1991). A *neo-evolutionary* model should capture the relations among the different functions (organized knowledge production, diffusion, and control) that operate in and on these networks. The functions have to be carried by the agents at the nodes, but one can no longer expect a one-to-one correspondence between functions and institutions because the functions are also based on the arrangements among the institutions (Etzkowitz & Leydesdorff, 2000). More important than the mere presence of agencies is the quality of their relations in a given configuration. Since the functions are performed by different agents and relations, one expects an uncertainty which can be measured as a probabilistic entropy. Systemic effects may occur that cannot be directly traced back to specific exchanges, but emerge more indirectly (Burt, 1995).

In this study, the different functions will be modeled as sub-dynamics of the system. The sub-dynamics can be expected to interact to varying degrees. The mutual information flowing between these sub-dynamics will be proposed as an indicator for the measurement of the synergy at the systems level.

The geographical distribution of industrial activities among regions is only one of the relevant dimensions of a configuration. Due to differences in the character of innovation processes, one can expect that geographical conditions have different effects on the various economic sectors such as manufacturing and knowledge-intensive services. We shall further distinguish between medium and high-tech using the classifications of the OECD (see Table 2 below). The division of labour among corporations of various sizes (e.g., the number of SMEs in a region) can be considered as a third determining factor (Storper, 1997). We use these three indicators (geography, technology, and firm size) and analyze their mutual information at various levels of the German system (states, regions) in order to test two hypotheses that we generated in a previous study using Dutch data (Leydesdorff *et al*., 2006):

1. medium-tech manufacturing can be considered as the drivers of the knowledge base of an economy more than high-tech;
2. knowledge-intensive services (KIS) tend to uncouple the knowledge base of an economy from its geographical location.

In summary: this paper introduces a way of assessing the quality of regional innovation systems by measuring the interaction and synergy between subsystems by means of an indicator based on entropy statistics (Jakulin & Bratko, 2004). The approach is applied to the various regions of Germany. The following section presents the conceptual basis of the study. Section 3 outlines the data and the spatial framework of the empirical analysis, and section 4 presents the method. Following the general assessment of the quality of regional innovation systems (section 5), we compare results for different sub-sectors of the economy, in particular high- and medium-tech manufacturing and knowledge-intensive services (section 6). Conclusions and policy implications are presented in the final section (section 7).

## 2. Theoretical background

Because innovation processes involve the generation *and* application of knowledge the quality of innovation systems is dependent on how the knowledge base is related to the network among the interacting agents (Foray, 2004). The network mainly provides an infrastructure to the innovation system: it facilitates and constrains exchanges of knowledge and resources. For a number of reasons such as the costs and effort involved in having face-to-face contact, a considerable part of these exchange relations is constrained geographically. The distribution of the technologies in a system, the industrial organization, and the geographical spread can be considered as relatively independent sources of variation. One can expect that these three sources of heterogeneity are reflected in the division of innovative labor.

It is important to note that the organization of the division of innovative labor does not necessarily require direct interaction, but can also be 'systemic' in nature, steered for example by market forces. For this reason, an analysis of the direct relationships of actors in regional innovation systems such as market relations and R&D cooperation may not provide a sufficient basis for assessing the working of the system. The geographical dimension first *positions* the agents involved; secondly, economic exchange *relations* can be expected among the agents at the nodes; and thirdly, the *dynamics* of knowledge-based innovations may upset the tendency towards equilibrium prevailing in economic exchange relations (Schumpeter [1939] 1964; Nelson & Winter, 1982).

The knowledge base of an economy can be considered as a trilateral interaction effect in addition to the bilateral interaction terms between each two of these subdynamics. However, this synergy at the level of the system has posed a problem for measurement. Our research question is whether one is able to operationalize an indicator of the emerging and therefore 'elusive' order of a knowledge-based economy (Skolnikoff, 1993) and then also to measure this order as a reduction of the uncertainty which prevails at the systems level (Carter, 1996; David & Foray, 2002).

The feedback loops between the knowledge infrastructure, the prevailing regime of a political economy, and the innovative dynamics in the market can be expected to change over time. Changes in these interactions drive cycles that can be longer-term than business cycles (Abernathy & Clark, 1985; Freeman & Perez, 1988; Schumpeter [1939] 1964). Whether or not, and to what extent, a knowledge-based economy has emerged from a specific configuration of relations remains an empirical question (Nelson, 1993; Storper, 1997). In short, the knowledge infrastructure of institutional relations (e.g., among universities, industries, and governments) can be considered as a necessary but not a sufficient condition for developing a knowledge-based economy. The intensity and the quality of the interactions is decisive for the characteristics of such a system.

Nations and regions can be expected to differ in combining the functional requirements of a knowledge-based economy. When a knowledge base is resulting from the synergy at the systems level, one can expect the system increasingly to 'self-organize' an additional feedback loop. This feedback may operate positively (that is, by reducing uncertainty in the relations) or negatively because, for example, it reinforces globalization in a previously more localized system. Etzkowitz & Leydesdorff (2000) called this additional feedback the operation of 'a network overlay' potentially emerging within a Triple Helix. In other words, the network of relations may turn into a configuration that can be productive, innovative, and flourishing, but not all networks can be expected to do so all the time.

For example, despite their productivity, innovativeness, and the density of relations networked in different dimensions (Biggiero, 1998), industrial districts and regions may suffer from deindustrialization because of the globalizing dynamics in the appropriation of the profits and the advantages of innovation (Beccatini *et al*., 2003). The *neo-institutional* perspective of social network analysis has provided us with a view of the (potentially changing) relations in the districts, but not on the dynamics. From this perspective, the emergence of a knowledge-based overlay to the system remains an unpredictable effect.

The *neo-evolutionary* model analyzes the Triple Helix dynamics in terms of how these relations operate: how much uncertainty is generated and/or reduced, at which level, and in which dimensions? One can expect an additional reduction of the uncertainty in the configuration if the overlay feeds *back* on the generation of uncertainty in the institutional relations. This decrease of uncertainty results from the configuration of relations and can no longer be attributed to the individual agents at the nodes or to specific relations.

The research question thus becomes, ‘To what degree is an emerging Triple Helix dynamics conducive to the development of specific regions and nations?’. Our data enable us to compare the results of an analysis of 438 districts (*Kreise*) in Germany with the conclusions from a similar study in the Netherlands (Leydesdorff *et al*., 2006). We use three proxies: (1) the geographical location points to the relevant governance structures (i.e., districts, regions, states); (2) the three-digit code of the industry as an indicator for the technological knowledge base; and (3) average firm-size as a measure of the organizational structure. The data enable us to cross-tabulate these three dimensions at the district level. In the Dutch study, one of us had obtained finer-grained data at the firm level. Nevertheless, some of the conclusions from the previous study can be tested against those emerging from the German data.

For example, we will be able to check the conclusion that the regional differences in the configurations are determined almost exclusively by high- and medium-tech *manufacturing*. The economic benefits of knowledge-intensive *services* are not located at the level of the regional innovation system but at the national level, while knowledge-based manufacturing tends to remain geographically embedded. Secondly, we can test our previous hypothesis that medium-tech manufacturing contributes more than high-tech production to the knowledge-based configuration. Corroboration of these two hypotheses has important implications for industrial development policies. Thirdly, we will compare the results for the whole of Germany with those for the former Eastern and Western parts of the country, and at the level of the Federal States (*Länder*).

## 3. Data

The employment and establishment data for this study were collected from the German Social Insurance Statistics (*Statistik der sozialversicherungspflichtig Beschäftigten*). These statistics are generated by the Federal Employment Office (*Bundesagentur für Arbeit*) (Fritsch & Brixy, 2004). In Germany, all public and private employers are required by law to register their employees with this office for enrollment in the social insurance and pension systems. In the case of composite (e.g., international) corporations with multiple locations, the data are collected at the level of the local establishments, and thus the geographical dimension is reflected in these data. However, employees who are not obliged by law to contribute to these insurance systems are (by definition) excluded from the statistics. These include, among others, civil servants, army personnel, the self-employed, and the unemployed.[3]

The statistics were made available to us at the NUTS-3 level of the Eurostat classification of regions (Eurostat, 2003). (NUTS is an abbreviation of *Nomenclature des Unités Territoriales Statistiques*.)[4] In the German Federal Republic the NUTS-3 level coincides with the district or *Kreis*. Eurostat (2003) distinguishes 440 of these districts in Germany. One of these is an unclassified category entitled 'extra region.' Two regions (Eisenach and Wartburg) have not always been distinguished in the German statistics and were merged for the purpose of this study.[5] Thus, we focus on 438 districts as the units of analysis. These districts are organized at the NUTS-2 level into 41 regions, which are called *Regierungsbezirke* in German. The NUTS-1 level is defined as the 16 Federal States or *Länder* that comprise the German Federal Republic. Figure 1 shows the organization of Germany into *Länder* and *Regierungsbezirke*, respectively. For information, the previously East-German part of the country is shaded in figure 1a.

[3] In manufacturing, the Social Insurance Statistics cover more than 90 percent of all employees. In the service sector, this share is about 80 percent. Coverage is relatively low in agriculture (less than 24 percent) and in the public sector (about 50 percent).

[4] The Nomenclature of Territorial Units for Statistics (NUTS) was established by Eurostat more than 25 years ago in order to provide a single uniform breakdown of territorial units for the production of regional statistics for the European Union; at http://europa.eu.int/comm/eurostat/ramon/nuts/introduction_regions_en.html

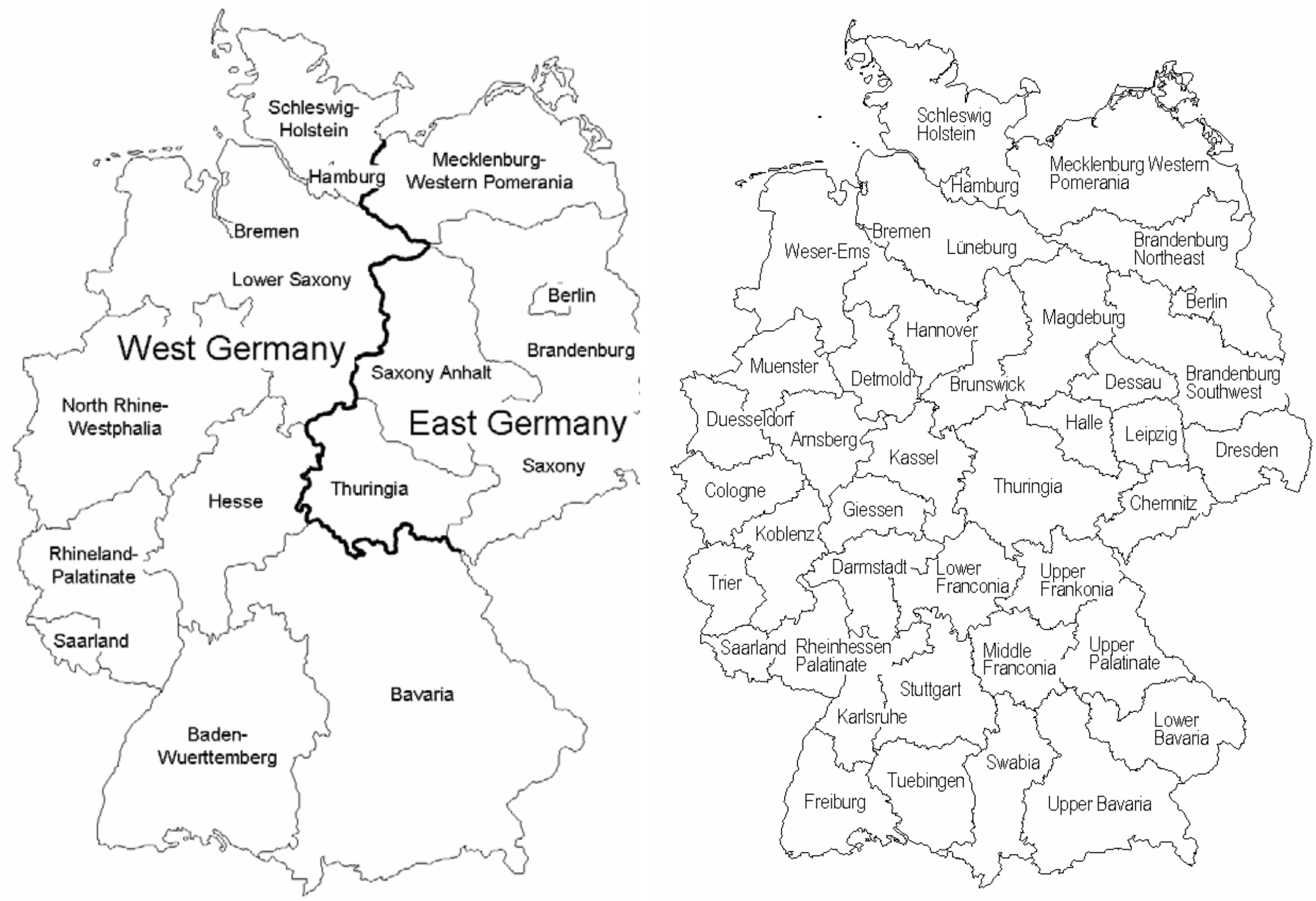

**Figure 1a and b:** The political administration of the German Federal Republic at the NUTS-1 (*Länder*) and NUTS-2 (*Regierungsbezirke*) levels, respectively.

For historical reasons, the cities of Berlin, Hamburg, and Bremen have been considered administratively as NUTS-1 categories (*Länder*). Berlin and Hamburg consist only of single districts, which are defined under the same name at the NUTS-3, NUTS-2, and NUTS-1 levels. Bremen is subdivided at the NUTS-3 level in two districts (Bremen and Bremerhaven). Other large cities, like Munich, Cologne, and Frankfurt, are defined as districts at the NUTS-3 level within their respective regions and states. For the purpose of this study, we decided to modify the data by adding Berlin as a district (at the NUTS-3 level) to Brandenburg – South-East (NUTS-2: DE42), Hamburg to the region Schleswig-Holstein (NUTS-2: DEF0), and the two districts Bremen and Bremerhaven to the region of Lüneburg (NUTS-2: DE93). At the NUTS-1 level Berlin will thus be considered as part of Brandenburg (DE4), and Hamburg as part of Schleswig-Holstein (DEF), while Bremen/Bremerhaven is assigned to Lower Saxony (DE9).

[5] The two districts 16056 (the city of Eisenach) and 16063 (the district Wartburg) have been distinguished administratively only since 1998. They are considered in this study as a single unit because of the comparisons over time that we envisage making in future studies.

The German Federal Office for Building and Regional Planning (*Bundesamt für Bauwesen und Raumordnung*; BBR, 2002) has attributed categories to these districts according to the settlement structure (Table 1). Not surprisingly, this classification is negatively correlated with the population density (population/area) of the districts at the level of Spearman's $\rho = -.67$ ($p < 0.01$; $N = 438$). Since some core cities of the urbanized regions (class 5) can be ranked higher in a hierarchy than the rural districts of the agglomerations (class 4), the category numbers do not necessarily indicate ranks in a spatial hierarchy. However, we will not use this classification as a rank order, but as a scheme generating a distribution in the geographical dimension.[6]

[6] The geographical units themselves are unique and, therefore, do not contain uncertainty.

**Table 1**: Classification of districts into district types

| Type of district | Classification number | Number of districts |
|---|---|---|
| *Agglomerations* | | |
| Core cities | 1 | 43 |
| Districts with high density | 2 | 44 |
| Districts with average density | 3 | 39 |
| Rural districts | 4 | 23 |
| *Urbanized regions* | | |
| Core cities | 5 | 29 |
| Districts with average density | 6 | 91 |
| Rural districts | 7 | 68 |
| *Rural areas* | | |
| Rural districts with relatively high density | 8 | 58 |
| Rural districts with relatively low density | 9 | 43 |
| | | N = 438 |

**Table 2**: Classification of high-tech and knowledge-intensive sectors according to the OECD and Eurostat. Source: Laafia, 2002: 7.

| *High-tech Manufacturing* | *Knowledge-intensive Sectors (KIS)* |
|---|---|
| **30** Manufacturing of office machinery and computers<br>**32** Manufacturing of radio, television and communication equipment and apparatus<br>**33** Manufacturing of medical precision and optical instruments, watches and clocks<br><br>*Medium-high-tech Manufacturing*<br><br>**24** Manufacture of chemicals and chemical products<br>**29** Manufacture of machinery and equipment n.e.c.<br>**31** Manufacture of electrical machinery and apparatus n.e.c.<br>**34** Manufacture of motor vehicles, trailers and semi-trailers<br>**35** Manufacturing of other transport equipment | **61** Water transport<br>**62** Air transport<br>**64** Post and telecommunications<br>**65** Financial intermediation, except insurance and pension funding<br>**66** Insurance and pension funding, except compulsory social security<br>**67** Activities auxiliary to financial intermediation<br>**70** Real estate activities<br>**71** Renting of machinery and equipment without operator and of personal and household goods<br>**72** Computer and related activities<br>**73** Research and development<br>**74** Other business activities<br>**80** Education<br>**85** Health and social work<br>**92** Recreational, cultural and sporting activities<br><br>Of these sectors, **64, 72** and **73** are considered *high-tech services*. |

In addition to information at the level of each district, such as the population and the size of the district, our data contain the numbers of establishments and employees in each district at the three-digit level of the NACE classification.[7] Since various sectors of the economy can be expected to use different technologies, the sector classifications can be used as a proxy for the technology (Pavitt, 1984). The OECD (2001, pp.137ff.) indicated the various sectors in terms of their knowledge intensity at the two-digit level of the NACE code as provided in Table 2. The 222 NACE categories present in our data are available at the three-digit level; these can be subsumed under 60 NACE categories at the two-digit level. We use the information at the three-digit level for the computation, but the two-digit level for making appropriate selections according to the definitions of the OECD and Eurostat.

**Table 3**: Distribution of company sizes in the German data.

| Number of employees | Frequency | Percent | Cumulative Percent |
|---|---|---|---|
| 1 | 4721 | 7.0 | 7.0 |
| 2 to 4 | 16117 | 24.0 | 31.0 |
| 5 to 9 | 17416 | 25.9 | 56.9 |
| 10 to 19 | 12690 | 18.9 | 75.7 |
| 20-49 | 9745 | 14.5 | 90.2 |
| 50-99 | 3501 | 5.2 | 95.4 |
| 100-199 | 1775 | 2.6 | 98.1 |
| 200-499 | 912 | 1.4 | 99.4 |
| 500-749 | 186 | .3 | 99.7 |
| 750-999 | 70 | .1 | 99.8 |
| > 1000 | 132 | .2 | 100.0 |
| | 67265 | 100.0 | |

For reasons of comparison with the Dutch study (Leydesdorff *et al.*, 2006), we used the same classification for the average establishment sizes (Table 3). Average firm size in terms of numbers of employees can be used as a proxy for the industrial organization (Pugh & Hickson, 1969; Pugh *et al.*, 1969; Blau & Schoenherr, 1971). The Dutch data included a category for firms with zero employment, but this category is not contained in the German statistics because self-employed persons are not obliged to contribute to the social insurance scheme. In summary, the maximum entropy of the system under study is determined by 222 NACE categories, 9 district types, and 11 size categories of establishments, that is, $H_{max}$ = $^2\log$ (222 * 9 * 11) = 14.42 bits of information (Theil, 1972; Leydesdorff, 1995).

[7] NACE stands for *Nomenclature générale des activités économiques dans les Communautés Européennes*. The NACE code can be translated into the so-called International Standard Industrial Classification.

## 4. Methods

According to Shannon's (1948) formula, the uncertainty in the distribution of a variable *x* ($\sum_x p_x$) can be measured as $H_x = -\sum_x p_x \, ^2\log p_x$. Aanalogously, $H_{xy}$ represents the uncertainty in the two-dimensional probability distribution (matrix) of *x* and *y* (that is, $H_{xy} = -\sum_x \sum_y p_{xy} \, ^2\log p_{xy}$). In the case of two dimensions, the uncertainty in the two interacting dimensions (*x* and *y*) is reduced with the mutual information or transmission $T_{xy}$. This can be formalized as follows:

$$T_{xy} = (H_x + H_y) - H_{xy} \qquad (1)$$

In the limiting case that the distributions *x* and *y* are completely independent, $T_{xy} = 0$ and $H_{xy} = H_x + H_y$. In all other cases $T_{xy} \geq 0$, and therefore $H_{xy} \leq H_x + H_y$ (Theil, 1972, at pp. 59f.).

Because of the sigma in the formulae, all information terms can be fully decomposed; if base two is used for the logarithm, all values are expressed in bits of information. Note that these measures are formal (probability) measures and thus independent of size or any other reference to the empirical systems under study. In general, two interacting systems (or variables) determine each other in their mutual information ($T_{xy}$) and condition each other in the remaining uncertainty on either side ($H_{x|y}$ and $H_{y|x}$, respectively).[8]

Abramson (1963: 129) derived from the Shannon formulas that the mutual information in three dimensions is:

$$T_{xyz} = H_x + H_y + H_z - H_{xy} - H_{xz} - H_{yz} + H_{xyz} \qquad (2)$$

While the bilateral relations between the variables reduce the uncertainty, the trilateral term in turn feeds back on this reduction, and therefore adds another term to the uncertainty. The layers alternate in terms of the sign. The configuration thus determines the net result in terms of the value of $T_{xyz}$. The potential reduction of the uncertainty at the systems level cannot be attributed to individual nodes or links. McGill (1954) proposed to use the term 'configurational information' for this mutual information in three (or more) dimensions, but with the opposite sign (Jakulin & Bratko, 2004).

The alterations can be generalized for more than three dimensions, but for reasons of parsimony we limit the discussion here to three dimensions. (Leydesdorff (2003) provides a graphical representation of the mutual information in a Triple Helix using Venn-diagrams.) As noted in the previous section, the three dimensions under study in this case will be geography, technology, and organization, and the measure will accordingly be indicated as the $T_{GTO}$. The value of $T_{GTO}$ measures the interrelatedness of the three sources and the fit of the relations between and among them. Because it is a measure of the *reduction* of the uncertainty, a better fit will be indicated with a more negative value. This overall reduction of the uncertainty can be considered as a result of the intensity and the productivity of innovative labor division in a broad sense.

[8] The transmission can be considered as an information-theoretical equivalent of the covariance as a measure of the covariation. The covariation is only a part of the total variation in each of the covarying dimensions. Unlike the covariance, the mutual information can be provided with an interpretation in the case of more than two dimensions and with a dynamic interpretation so that a co-evolution can also be measured (Leydesdorff, 1995).

Assuming that a division of labor can yield efficiency gains, one would expect that regions with a distinctive profile would be more productive than regions with a lower level of division of labor. However, the indicator does *not* measure the innovative activity or economic output of a system (Carter, 1996). It measures only the structural conditions in the system for innovative activities, and thus specifies an expectation. Regions with a high potential for innovative activity can be expected to organize more innovative resources than regions with lower values of the indicator.

## 5. Results

Figure 2 shows the results of the computations using the specification of the data and methods above for the whole of Germany, aggregated at both the NUTS-1 level (*Länder*) and the NUTS-2 level of *Regierungsbezirke*. The left-hand figure (at the NUTS-1 level) exhibits the different dynamics in the former eastern and western parts of the country. This is not surprising given the need for radical reorganization of the East German innovation system after re-unification. The socialist type of innovation regime that existed in the former German Democratic Republic until the fall of the Iron Curtain in 1989 was so different from a market-based system that the transition between these two regime types can be expected to take considerable time (Fritsch & Werker, 1999). It is, however, remarkable that the weakest knowledge base is found for the West-German region of Saarland, and that the East-German

NUTS-1 level unit Saxony seems already to perform better on this indicator than the West-German state of Schleswig-Holstein (see Table 4).

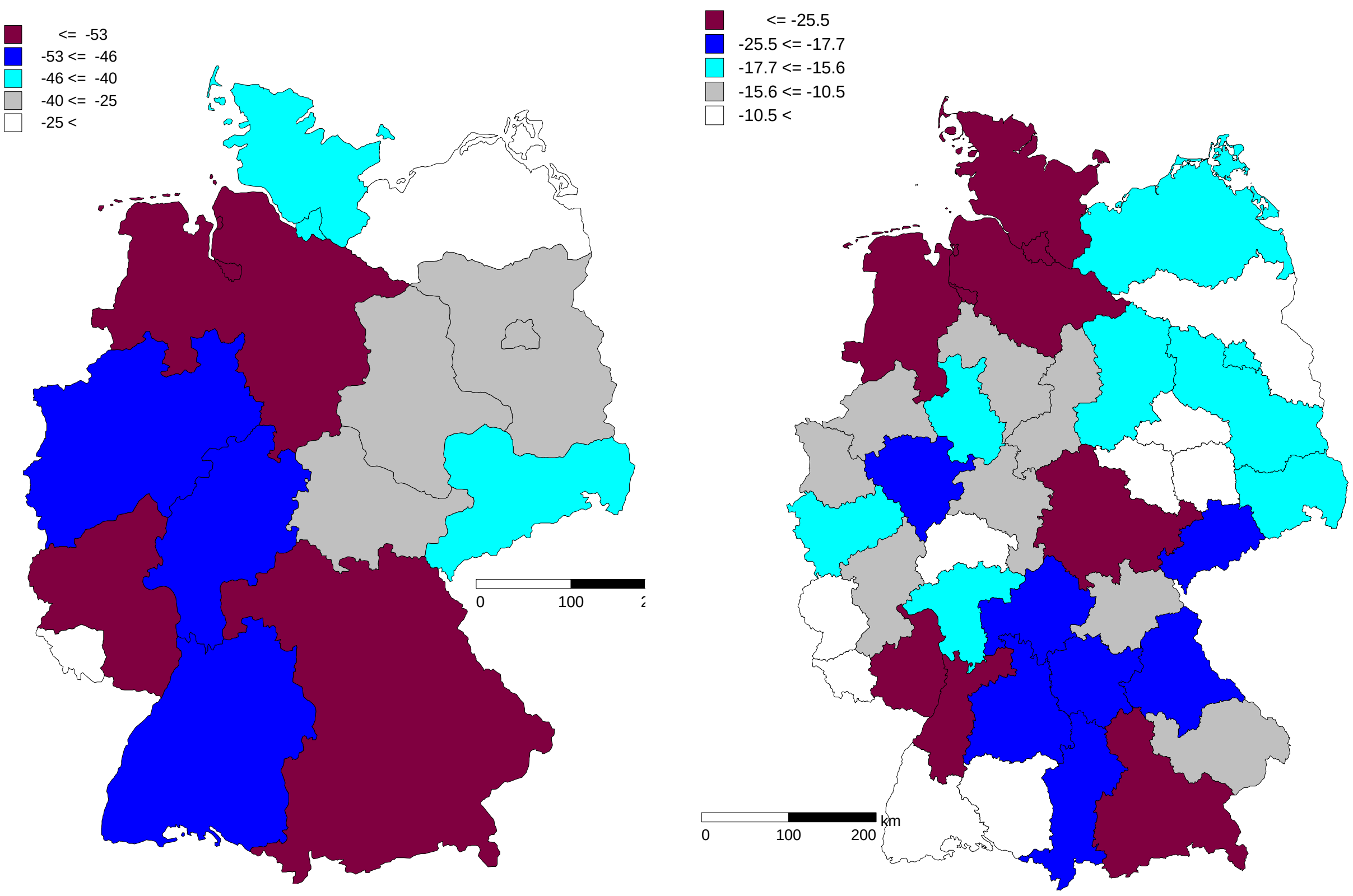

**Figure 2a**: The mutual information in three dimensions ($T_{GTO}$) at the NUTS-1 level

**Figure 2b:** Idem at the NUTS-2 level.

The right-hand figure (at the NUTS-2 level) shows a more differentiated picture. It highlights high levels of interaction and knowledge flows in the metropolitan areas of Munich and Hamburg. In addition to Saarland, the regions of Trier, Gießen, Freiburg, and Tübingen have similarly low values in the western part of Germany. In East Germany the weakest knowledge base is found in the rather sparsely populated Northeast Brandenburg, as well as in the neighbouring regions of Leipzig, Halle, and Dessau. While the region of Northeast Brandenburg is nearly devoid of R&D resources, the regions of Leipzig, Halle, and Dessau are old industrialized areas in which the need for radical change has been particularly strong.

**Table 4.** The mutual information in three dimensions statistically decomposed at the NUTS-1 level of the German states (*Länder*).

| NUTS 1 (Länder) | $T_{GTO}$ in mbits before normalization | $\Delta T_{GTO}$ (= $n_i$ * $T_i$ /N) in mbits of information | $n_i$ |
|---|---|---|---|
| Baden-Württemberg | -474.91 | -47.71 | 44 |
| Bavaria | -412.48 | -90.41 | 96 |
| Brandenburg | -583.86 | -25.33 | 19 |
| Hesse | -778.93 | -46.24 | 26 |
| Mecklenburg-Western Pomerania | -430.28 | -17.68 | 18 |
| Lower Saxony | -632.35 | -69.30 | 48 |
| North Rhine-Westphalia | -404.10 | -49.82 | 54 |
| Rhineland-Palatinate | -647.76 | -53.24 | 36 |
| Saarland | -639.67 | -8.76 | 6 |
| Saxony | -649.83 | -43.03 | 29 |
| Saxony-Anhalt | -600.14 | -32.88 | 24 |
| Schleswig-Holstein | -1102.75 | -40.28 | 16 |
| Thuringia | -619.48 | -31.12 | 22 |
| Germany | -180.08 | -180.08 | 438 |

The two pictures are based on different normalizations because the contributions of regions are weighted in terms of the number of districts at the respective level of aggregation, but with reference to the values for Germany as a whole. The districts (at the NUTS-3 level) are our units of analysis, while the NUTS-2 level and the NUTS-1 level are levels of aggregation. In addition to the aggregation, however, one would also expect interaction effects at the national level. Given the different normalizations at the NUTS-2 or NUTS-1 level, the two representations cannot be compared directly in terms of the absolute values of the indicator.

For example, the states of Schleswig-Holstein (formerly part of Western Germany) and Mecklenburg-Western Pomerania (formerly part of Eastern Germany) are defined as units both at the NUTS-1 and the NUTS-2 level, and thus these units are comparatively large when compared with other regions in the right-hand picture, while the same values are compared with the other states in the left-hand figure. In both figures, the contribution of each part of the country is normalized with reference to the country (that is, Germany; N = 438) as the baseline using $\Delta T_{GTO} = n_i * T_i / N$. Since the districts are our units of analysis, $n_i$ in this formula stands for the number of districts in the unit under study and $T_i$ for the mutual information in the three dimensions – Geography, Technology, and Organization – at this level of aggregation. Table 4 provides these values for the *Länder* (that is, at the NUTS-1 level). At the NUTS-2 level the values for $n_i$ are lower except for those NUTS-1 level units that are not further decomposed (e.g., Schleswig-Holstein and Mecklenburg-Western Pomerania). The values at the NUTS-2 level are provided in the Appendix.

Unlike our previous results for the Netherlands, the value of the indicator for Germany as a whole is less negative then the sum of the values for the *Länder*. This means that there is configurational synergy at the local levels of NUTS-1 and NUTS-2 that is no longer apparent when the distributions are aggregated at the national level. The negative entropy is a local attribute. When decomposed, the additional synergy is generated mainly by the mutual information between the NACE-codes and the size categories of the business. The type of district (whether it is rural or urban) has less influence on the potential synergy than the interplay between the organizational format and the technological structure of the industry.

Should perhaps the Netherlands as a country rather be compared with the separate states of the Federal Republic? We are not able to answer this question on the basis of the data because the results for the Netherlands were based on micro-data, and we did not use a characterization of the districts in terms of whether they were rural or urban, but only on the basis of postal codes. However, Figure 3 shows the results of limiting the analysis to Bavaria as an example of such a lower-level decomposition at the level of a state.

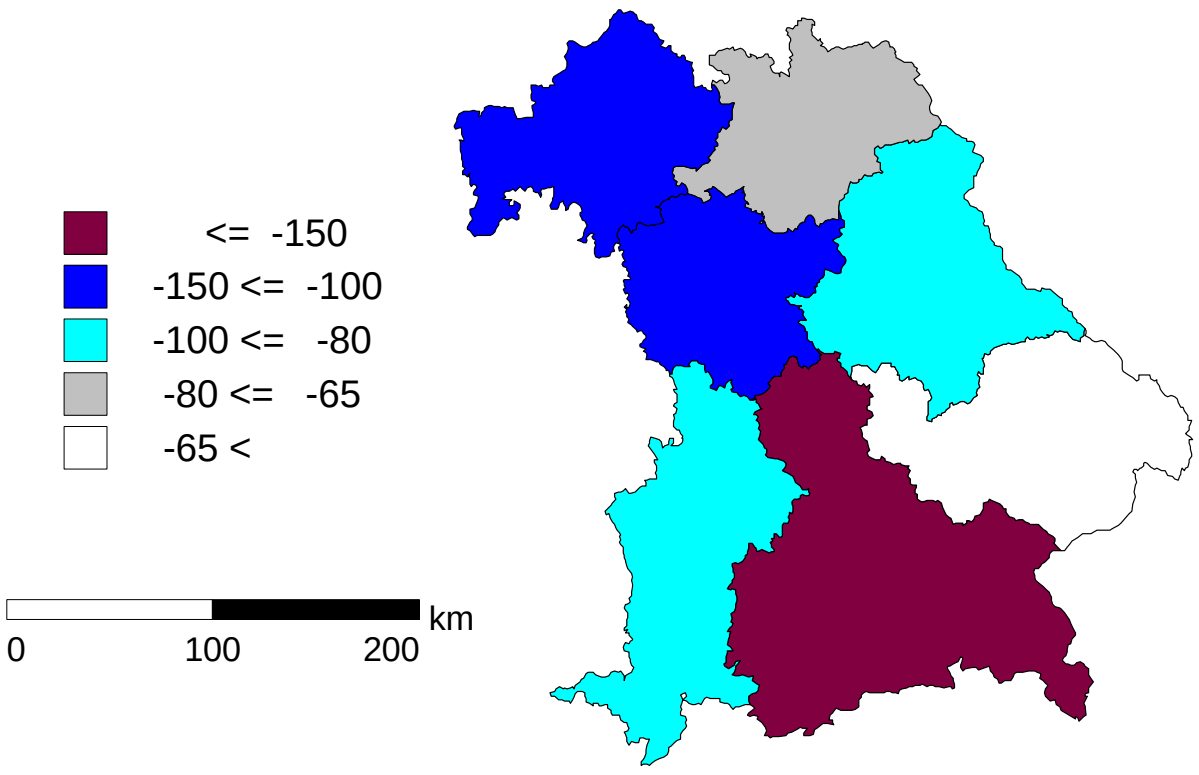

**Figure 3**: The mutual information in three dimensions normalized for the 7 regions (*Regierungsbezirke*) of Bavaria.

The picture that we obtain in our analysis for Bavaria (Figure 3) corresponds particularly well with the commonly assumed quality of the innovation system in the different regions of this state. Unfortunately, information about the quality of regional innovation systems is hitherto available only for certain regions, often in the form of case studies that do not allow for a systematic interregional assessment and comparison with our indicator. According to such case studies the innovation system of the Munich region has been said to be highly productive in recent times (Sternberg & Tamasy, 1999; Krauss & Wolff, 2002), while the Ruhr area and some of the East-German regions are lagging behind, with the exception of Saxony and Thuringia (Fritsch & Slavtchev, 2006). This corresponds with our findings at the NUTS-2 level (figure 2b).

The pronounced position of the metropolis of Munich in Figure 5 contrasts with the lowest rank for the innovation system in the region of Lower Bavaria (east of Munich), which has a reputation for being characterized by a relatively low level of dynamics (Fritsch & Slavtchev, 2006). This region, as well as Upper Palatinate and Upper Franconia, was located on the border with the Czech Republic and the former German Democratic Republic, respectively. The Iron Curtain that divided Eastern and Western Europe for a long period of time may have left a longer lasting imprint on these regions. Upper Franconia in particular is rather peripheral and at some distance from any larger centre. The region of Middle Franconia contains Nuremberg, the second largest city of the state of Bavaria, while the region of Lower Franconia is adjacent to the dynamic Frankfurt area. According to our calculations, both regions are in the second highest category. One may assume that Swabia draws some benefits from its geographical closeness to Munich and to Nuremberg so that it maintains a middle-range position.

## 6. The sectorial decomposition in terms of the knowledge base

As noted earlier, one main purpose of this article was to test two hypotheses which emerged from a previous study of the knowledge base of the Dutch economy. These two results of the study for the Netherlands were:

1. Medium-tech manufacturing generates more synergy in a geographical unit than high-tech establishments;
2. Knowledge-intensive services decouple the knowledge base from its geographical location, while high- and medium-tech manufacturing remain geographically embedded.

The interpretation of these two findings could be that the sectors assume different roles in the division of innovative labor. Medium-tech businesses can be expected to focus on maintaining absorptive capacity, so that knowledge and technologies developed elsewhere can be understood more easily and adapted to local circumstances (Cohen & Levinthal, 1989). From the perspective of the organization of technological knowledge, high-tech manufacturing may be focused on (internal) production within the transnational corporation, take place as spin-offs of research institutions, and involve global markets more than local environments. From an industrial perspective, one could even assume that medium-tech manufacturing functions as a seedbed for high-tech production.

The knowledge-intensive services can be expected to decouple from the geographical location more easily than manufacturing because these services can be offered across regional boundaries, for example, by using communication media or by traveling of the consultants. It is not uncommon for knowledge-intensive services to be offered on the site of the customer by someone brought in from elsewhere. Unlike manufacturing, knowledge-intensive services can be offered throughout the country and abroad without necessary links to local production facilities like factories. Thus, the geographical location can be chosen by these firms on grounds different from the local configuration in terms of the Triple Helix dimensions. Note that this reasoning can be expected to hold less for knowledge-intensive sectors which are also high-tech, because in this case a local production component (e.g., R&D laboratories) may be needed for support.

**Table 5**: The distribution of records and establishments in the database across sectors given different selections using NACE codes.

| 2002 | Number of records | Number of establishments | Number of employees | NACE categories |
|---|---|---|---|---|
| All sectors | 67,265 | 2,119,028 | 27,596,100 | |
| High-tech manufacturing | 2,606<br>*3.9%* | 23,912<br>*1.1%* | 663,210<br>*2.4%* | 30, 32, 33 |
| Medium-tech manufacturing | 7,127<br>*10.6%* | 39,281<br>*1.9%* | 2,820,436<br>*10.2%* | 24, 29, 31, 34, 35 |
| Knowledge-intensive services | 17,271<br>*25.7%* | 703,817<br>*33.2%* | 9,619,657<br>*34.9%* | 61, 62, 64, 65, 66, 67, 70, 71, 72, 73, 74, 80, 85, 92 |
| High-tech services | 3,097<br>*4.6%* | 45,485<br>*2.1%* | 817,305<br>*3.0%* | 64, 72, 73 |

Table 5 compares the relevant numbers of units (e.g., establishments) with the respective NACE categories as provided in Table 2 above. While in the Netherlands 51.3% of the establishments were involved in knowledge-intensive services, this percentage is only 33.2% for Germany. Note that in accordance with the OECD/Eurostat classifications the high-tech services are considered as a subclass of knowledge-intensive services, while high- and medium-tech manufacturing are considered as two separate classes. The ratio between high- and medium-tech manufacturing establishments is 23,912/39,281 = 0.61 for Germany versus 4,126/11,712 = 0.35 for the Netherlands.[9] This confirms that Germany is relatively more high-tech in manufacturing, while less knowledge-intensive in the service sectors.

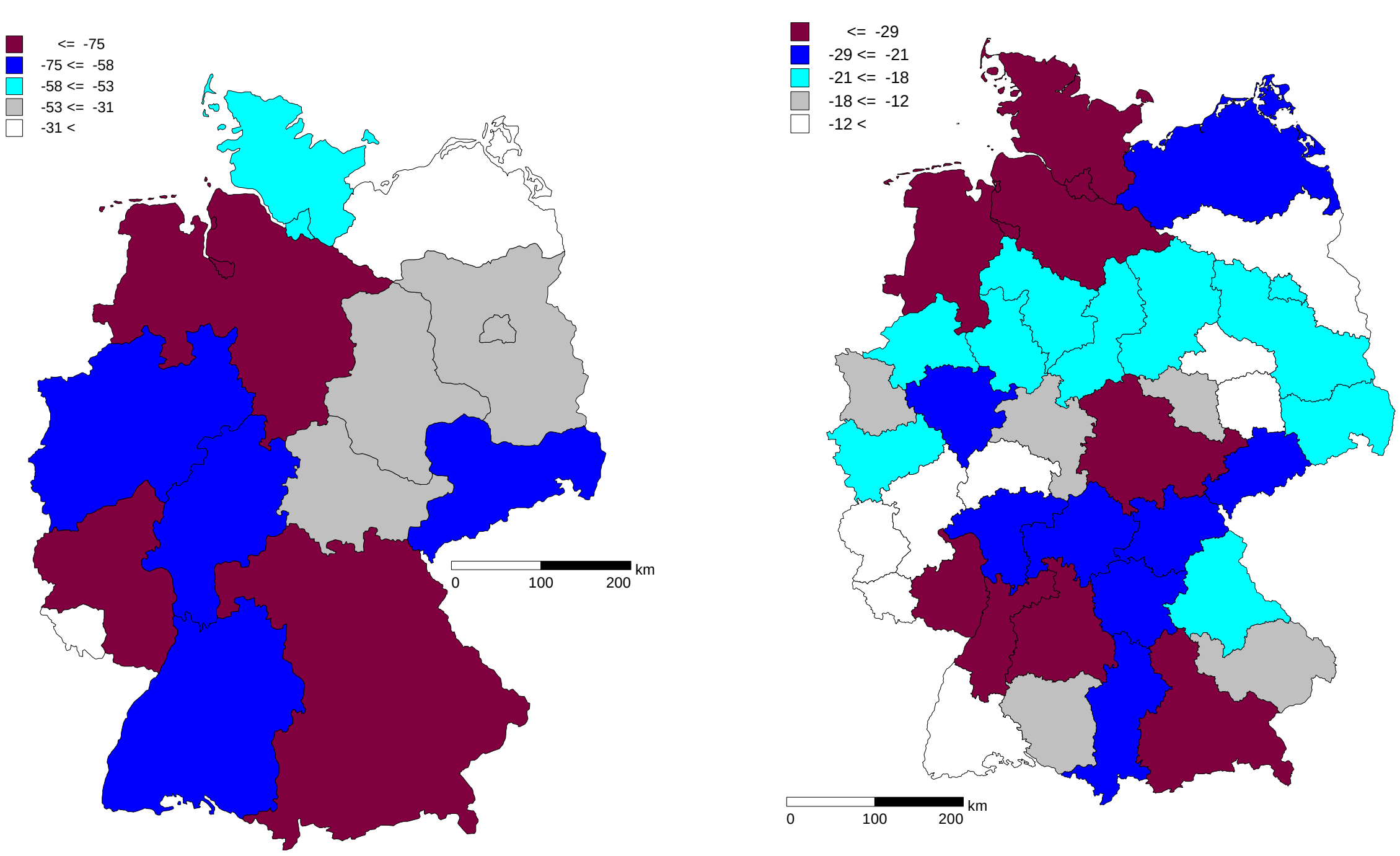

**Figure 4a and b**: Medium-tech manufacturing generates the knowledge base in Germany.

[9] Unfortunately, the Dutch data do not allow us to make this comparison in terms of the numbers of employees in the different sectors because the respondents indicated only the size categories. However, more than 71.8% of the high-tech manufacturing firms have a size smaller than five employees, while the figure is only 61.2% for medium-tech frims. For Germany, the ratio of 663,210 employees in high-tech versus 2,820,436 in medium-tech manufacturing is 0.23.

Figure 4 shows the results when the analysis is limited to medium-tech manufacturing, that is, approximately 1.9 % of the total number of establishments included in the data. The pictures are virtually similar to those in Figure 2, which were based on 100% of the data. As expected, some regions are ranked higher when we focus on this selection, but the pattern is broadly the same. In other words, the quality of the regional innovation system is more or less completely determined by medium-tech manufacturing. The tables in the Appendix show us that high-tech manufacturing reduces the (negative) configurational information more often than not, while the effects of medium-tech always make the configurational information more negative. Therefore, the configuration of medium-tech manufacturing can be considered a better indicator of the knowledge-based economy than that of high-tech manufacturing.

The relation between high- and medium-tech manufacturing thus exhibits the patterns that were predicted on the basis of the study for the Netherlands. With the exception of Mecklenburg-Western Pomerania, the configurational information gain is always larger for medium-tech than for high-tech manufacturing at the NUTS-1 level. At the NUTS-2 level high-tech manufacturing provides a decrease of the configurational information in 26 of the 38 regions, while medium-tech manufacturing always increases the reduction of the prevailing uncertainty. These effects are not specific to the western or eastern parts of Germany.

The tables in the Appendix further reveal that knowledge-intensive services have the effect of making the configurational information less pronounced. With the exception of Hesse, all states (at the NUTS-1 level) are further coupled geographically when from the knowledge-intensive services only the high-tech ones are selected. For Mecklenburg-Western Pomerania and Saxony-Anhalt (both parts of the former GDR) these couplings even become a contribution to the configurational information. At the NUTS-2 level, such a contribution of high-tech services is also found in certain West-German regions (e.g., Koblenz), but this effect is more pronounced in East-German regions like Magdeburg.

**Table 6**: Variances in the ΔT across knowledge-based sectors of the economy at the NUTS-1 and NUTS-2 levels

| | NUTS-1 | NUTS-2 |
|---|---|---|
| All sectors | 458.277 | 83.763 |
| knowledge intensive services | 256.176 | 57.624 |
| high-tech services | 357.369 | 75.007 |
| medium-tech manufacturing | 1133.959 | 161.110 |
| high-tech manufacturing | 744.390 | 91.898 |
| high & medium-tech manufacturing | 1058.008 | 106.112 |
| Number of regions | 13 | 38 |

Table 6 expresses these conclusions in quantitative terms by using the respective variances as a measure of the synergetic effects. The variances clearly show that the levels of regional difference are considerably smaller for high-tech manufacturing than for medium-tech manufacturing, but as we have seen above both have a structuring effect on the economy. The knowledge-intensive services do not contribute to structuring the knowledge-based economy differentially among regions, but this is less the case at the high-tech end of these services. As noted, this latter effect is enhanced in less-developed regions like Eastern Germany.

## 7. Conclusions

Our analyses indicate that the federal structure of Germany makes the states (*Länder*) probably a more important unit of analysis in studying the knowledge-based economy than the Federal Republic as a whole. Using the mutual information in three dimensions as an indicator, we were able to reproduce the former division of the country between East and West at the NUTS-1 level of the states, but at the lower (NUTS-2) level of regions (*Regierungsbezirke*) the picture has in the meantime become more complex. We indicated several dynamics that are different for East and West—for example, the function of high-tech knowledge-intensive services seems to function more locally in the Eastern part of Germany—but we also note a further merging of the two parts of the country in terms of it's knowledge-based dynamics. One should keep in mind that globalization and the emergence of a knowledge-based economy itself were partly effects as well as cause of the demise of the Soviet Union and the subsequent integration of Germany (Leydesdorff, 2000).

The Netherlands has been a nation state since the time of Napoleon. Nevertheless, we found as much variation in the Netherlands as in Germany in terms of the mutual information among technology indicators, business indicators, and geographical locations. One technical reason for this is, perhaps, that the data for the Netherlands were finer-grained than for Germany. From the perspective of evolutionary theory, however, an emerging feedback operating as an additional selection mechanism can be expected to produce a skewed distribution in the underlying variation. Thus, one can expect that some regions will tend to become more knowledge-based in their economy than others despite efforts by national and regional governments to redistribute resources proportionately (Danell & Persson, 2003). In an increasingly globalized environment, mechanisms other than political control thus become more important than traditional policies (Bathelt, 2003; Cooke & Leydesdorff, 2005). The non-equilibrium dynamics of the knowledge-based economy can be expected to counteract the equilibrium-seeking mechanisms of the market and the quasi-equilibrium of redistribution by institutional policies (Aoki, 2001).

We have argued that the dynamics of technological innovation can be expected to add a third sub-dynamics to the political dynamics of institutionalization and regulation, and the equilibrating forces of the market (Leydesdorff & Meyer, 2003). This interaction among three sub-dynamics can be captured by using the evolutionary version of the Triple Helix model and can then be measured using the mutual information among these sub-dynamics. The mutual information in three dimensions can also be considered as configurational information (McGill, 1954; Jakulin & Bratko, 2004). In other words, we suggest measuring the knowledge base of an economy as a local configuration in which the uncertainty can be reduced because of a fit at the systems level.

The results of our measurements using this indicator confirmed the hypothesis that the quality of a regional innovation system is determined almost entirely by medium- and high-tech manufacturing. Indeed, the contribution of medium-tech manufacturing to the configuration can be used as a predictor of the properties of the innovation system in a given region. High-tech manufacturing adds to the pattern, but the size of high-tech manufacturing is small and the relative effect is also small. Because high-tech industry is so thinly spread across the country, its distribution may be more pronounced than for larger sectors.

Knowledge-intensive services seem to be largely decoupled from the configuration within a regional or local economy. They contribute negatively to the knowledge-based configuration because of the inherent capacity of the service providers to deliver these services outside the region. Thus, a locality can be chosen on the basis of considerations other than those relevant for the generation of a knowledge-based economy in the region. For example, the proximity of a well-connected airport (or train station) may be a major factor in the choice of a location.

This conclusion of the globalizing effect of knowledge-intensive services holds true for all regions both in the Netherlands and in Germany. However, the high-tech component of the knowledge-intensive services sometimes exhibits a coupling with the regional economy. This effect is particularly strong in some of the formerly East-German regions (e.g., Dessau, Magdeburg, Mecklenburg, and Western Pomerania). Given the prevailing pattern in more developed parts of the economy, this effect may disappear in the longer run because it may be specific to the developmental stage of the economy in these parts of Eastern Germany.

In summary, let us repeat the policy implications of this study: In many countries innovation policies have focused on the high-tech sector. According to our findings, medium-tech industry is at least as important for the local quality of the knowledge-based economy. This suggests that a strong focus on high-tech is not justified and that development policies should also account for other sectors, particularly the medium-tech industries. Insofar as one attracts knowledge-intensive services, however, these services should be stimulated at the high-tech end. Unlike knowledge-intensive services, high-tech services seem to have reinforcing effects on the configurational information.

Knowledge-intensive services can be important for generating employment, but one cannot expect these sectors to contribute significantly to the knowledge base of a regional economy. In addition to providing potentially high-quality employment, knowledge-intensive service providers that operate at an inter-regional level may be an important medium for knowledge spillovers across regions. When innovation policies are developed at the regional level, measures can more easily reflect the specific configurations of industrial structure and knowledge infrastructure in a region (Fritsch & Stephan, 2005). The synergetic effects of a knowledge-based economy vary among both regions and sectors.

## Acknowledgement

The authors are grateful to Pamela Mueller and Antje Weyh for their support in preparing the data for this study.

**Appendix 1**: Decomposition in terms of medium- and high-tech manufacturing versus high-tech and knowledge-intensive services

Table A1: T in mbits and change of T-values at the NUTS-1 level

| | All sectors | Manufacturing | | | | | | Services | | | |
|---|---|---|---|---|---|---|---|---|---|---|---|
| | | High-tech | Change (%) | Medium-tech | Change (%) | High- and medium-tech | Change (%) | Knowledge intensive | Change (%) | High-tech | Change (%) |
| Germany | -180.08 | -163.337 | -9.3 | -202.135 | 12.2 | -199.013 | 10.5 | -161.912 | -10.1 | -164.037 | -18.8 |
| Baden-Württemberg | -47.71 | -52.672 | 10.4 | -62.893 | 31.8 | -63.281 | 32.6 | -41.168 | -13.7 | -46.155 | -3.3 |
| Bavaria | -90.41 | -110.363 | 22.1 | -142.965 | 58.1 | -137.331 | 51.9 | -67.983 | -24.8 | -79.023 | -12.9 |
| Brandenburg | -25.33 | -20.696 | -18.3 | -30.564 | 20.7 | -29.213 | 15.3 | -21.06 | -16.8 | -22.447 | -11.4 |
| Hessen | -46.24 | -44.662 | -3.4 | -58.114 | 25.7 | -56.669 | 22.6 | -36.101 | -21.9 | -35.146 | -24.0 |
| Mecklenburg-Western Pomerania | -17.68 | -10.692 | -39.5 | -22.514 | 27.3 | -22.71 | 28.4 | -16.944 | -4.2 | -19.429 | 9.9 |
| Lower Saxony | -69.3 | -72.053 | 4.0 | -94.14 | 35.8 | -94.219 | 36 | -54.223 | -21.8 | -66.376 | -4.2 |
| North Rhine-Westphalia | -49.82 | -46.531 | -6.6 | -63.765 | 28.0 | -61.668 | 23.8 | -38.333 | -23.1 | -43.291 | -13.1 |
| Rhineland-Palatinate | -53.24 | -48.53 | -8.8 | -75.113 | 41.1 | -72.715 | 36.6 | -44.538 | -16.3 | -50.665 | -4.8 |
| Saarland | -8.76 | -9.123 | 4.1 | -8.792 | 0.4 | -9.721 | 10.9 | -7.951 | -9.3 | -8.712 | -0.5 |
| Saxony | -43.03 | -40.379 | -6.2 | -58.074 | 35.0 | -56.217 | 30.7 | -31.644 | -26.5 | -36.931 | -14.2 |
| Saxony-Anhalt | -32.88 | -22.845 | -30.5 | -46.793 | 42.3 | -43.659 | 32.8 | -26.831 | -18.4 | -33.386 | 1.5 |
| Schleswig-Holstein | -40.28 | -41.331 | 2.6 | -53.345 | 32.4 | -54.838 | 36.1 | -33.144 | -17.7 | -34.15 | -15.2 |
| Thuringia | -31.12 | -25.059 | -19.5 | -45.461 | 46.1 | -43.412 | 39.5 | -22.593 | -27.4 | -29.887 | -4.0 |

Table A2: T in mbits and change of T-values at the NUTS-2 level

| | All sectors | Manufacturing | | | | | | Services | | | |
|---|---|---|---|---|---|---|---|---|---|---|---|
| | | High-tech | Change (%) | Medium-tech | Change (%) | High- and medium-tech | Change (%) | Knowledge intensive | Change (%) | High-tech | Change (%) |
| Germany | -180.08 | -163.337 | -9.3 | -202.135 | 12.2 | -199.013 | 10.5 | -161.912 | -10.1 | -164.037 | -8.9 |
| Arnsberg | -17.93 | -18.711 | 4.4 | -23.859 | 33.1 | -23.822 | 32.9 | -11.947 | -33.4 | -14.806 | -17.4 |
| Brandenburg – Northeast | -5.52 | -4.042 | -26.8 | -5.878 | 6.5 | -5.607 | 1.6 | -4.801 | -13 | -5.25 | -4.9 |
| Brandenburg – Southwest | -17.62 | -15.759 | -10.6 | -19.135 | 8.6 | -19.18 | 8.9 | -15.036 | -14.7 | -16.32 | -7.4 |
| Brunswick | -13.81 | -19.051 | 38.0 | -18.087 | 31.0 | -19.649 | 42.3 | -10.506 | -23.9 | -15.534 | 12.5 |
| Chemnitz | -21.92 | -15.495 | -29.3 | -26.262 | 19.8 | -25.845 | 17.9 | -17.426 | -20.5 | -19.132 | -12.7 |
| Darmstadt | -17.06 | -16.374 | -4.0 | -21.064 | 23.5 | -20.881 | 22.4 | -13.578 | -20.4 | -12.381 | -27.4 |
| Dessau | -5.34 | -1.427 | -73.3 | -8.109 | 51.9 | -7.713 | 44.4 | -4.551 | -14.8 | -6.017 | 12.7 |
| Detmold | -15.78 | -11.895 | -24.6 | -17.794 | 12.8 | -17.389 | 10.2 | -13.608 | -13.7 | -15.21 | -3.6 |
| Dresden | -16.61 | -15.325 | -7.7 | -19.385 | 16.7 | -20.062 | 20.8 | -12.858 | -22.6 | -15.162 | -8.7 |
| Duesseldorf | -12.63 | -12.134 | -3.9 | -15.055 | 19.2 | -14.968 | 18.5 | -10.713 | -15.2 | -13.547 | 7.3 |
| Freiburg | -8.58 | -8.754 | 2.0 | -8.983 | 4.7 | -9.856 | 14.9 | -7.621 | -11.1 | -9.094 | 6.0 |
| Giessen | -3.67 | -2.806 | -23.5 | -4.582 | 24.9 | -4.342 | 18.3 | -3.032 | -17.4 | -3.204 | -12.7 |
| Halle | -10.31 | -4.612 | -55.3 | -12.306 | 19.4 | -11.307 | 9.6 | -8.788 | -14.8 | -4.795 | -53.5 |
| Hanover | -15.58 | -10.946 | -29.7 | -18.66 | 19.8 | -18.49 | 18.7 | -12.162 | -21.9 | -9.756 | -37.4 |
| Karlsruhe | -25.51 | -26.348 | 3.3 | -28.749 | 12.7 | -30.567 | 19.8 | -22.15 | -13.2 | -24.036 | -5.8 |
| Kassel | -13.71 | -5.462 | -60.2 | -13.975 | 1.9 | -13.681 | -0.2 | -10.933 | -20.2 | -11.801 | -13.9 |
| Koblenz | -11.08 | -9.439 | -14.8 | -11.302 | 2.0 | -11.744 | 6 | -9.761 | -11.9 | -11.991 | 8.2 |
| Cologne | -16.52 | -11.029 | -33.2 | -19.042 | 15.3 | -18.708 | 13.3 | -14.123 | -14.5 | -14.717 | -10.9 |
| Leipzig | -9.2 | -6.301 | -31.5 | -10.196 | 10.8 | -9.932 | 8 | -6.958 | -24.4 | -8.593 | -6.6 |
| Lüneburg | -29.69 | -19.320 | -34.9 | -38.109 | 28.4 | -36.888 | 24.2 | -26.414 | -11 | -28.161 | -5.1 |
| Magdeburg | -17.28 | -14.664 | -15.1 | -17.65 | 2.1 | -17.705 | 2.5 | -15.558 | -10 | -20.052 | 16.0 |
| Mecklenburg-Western Pomerania | -17.68 | -10.692 | -39.5 | -22.514 | 27.3 | -22.71 | 28.4 | -16.944 | -4.2 | -19.429 | 9.9 |
| Middle Franconia | -23.4 | -23.846 | 1.9 | -26.82 | 14.6 | -27.4 | 17.1 | -18.897 | -19.2 | -16.221 | -30.7 |
| Muenster | -13.84 | -13.632 | -1.5 | -17.577 | 27.0 | -17.551 | 26.8 | -12.35 | -10.8 | -11.838 | -14.5 |
| Lower Bavaria | -11.2 | -9.780 | -12.7 | -15.275 | 36.4 | -15.228 | 36 | -7.598 | -32.2 | -7.634 | -31.8 |
| Upper Bavaria | -41.88 | -40.463 | -3.4 | -54.101 | 29.2 | -53.566 | 27.9 | -34.495 | -17.6 | -34.719 | -17.1 |
| Upper Frankonia | -14.16 | -16.318 | 15.2 | -20.542 | 45.1 | -20.387 | 44 | -10.643 | -24.8 | -11.732 | -17.1 |
| Upper Palatinate | -17.92 | -15.371 | -14.2 | -20.145 | 12.4 | -20.367 | 13.6 | -14.145 | -21.1 | -15.118 | -15.6 |
| Rheinhessen-Palatinate | -33.07 | -34.937 | 5.6 | -49.102 | 48.5 | -48.831 | 47.6 | -28.737 | -13.1 | -34.158 | 3.3 |
| Saarland | -8.76 | -9.123 | 4.1 | -8.792 | 0.4 | -9.721 | 10.9 | -7.951 | -9.3 | -8.712 | -0.5 |
| Schleswig-Holstein | -40.28 | -41.331 | 2.6 | -53.345 | 32.4 | -54.838 | 36.1 | -33.144 | -17.7 | -34.15 | -15.2 |
| Swabia | -19.04 | -22.383 | 17.6 | -26.878 | 41.2 | -27.671 | 45.3 | -13.094 | -31.2 | -16.286 | -14.5 |
| Stuttgart | -23.61 | -24.511 | 3.8 | -28.782 | 21.9 | -29.855 | 26.4 | -20.072 | -15 | -23.589 | -0.1 |
| Thuringia | -31.12 | -25.059 | -19.5 | -45.461 | 46.1 | -43.412 | 39.5 | -22.593 | -27.4 | -29.887 | -4.0 |
| Trier | -4.78 | -1.416 | -70.4 | -5.737 | 20.0 | -5.256 | 9.9 | -4.568 | -4.5 | -3.392 | -29.0 |
| Tuebingen | -10.47 | -12.039 | 15.0 | -12.639 | 20.7 | -13.375 | 27.7 | -9.12 | -12.9 | -6.735 | -35.7 |
| Lower Franconia | -22.94 | -14.503 | -36.8 | -26.163 | 14.0 | -26.572 | 15.8 | -18.563 | -19.1 | -19.201 | -16.3 |
| Weser-Ems | -26.72 | -21.168 | -20.8 | -34.547 | 29.3 | -33.875 | 26.8 | -21.611 | -19.1 | -28.17 | 5.4 |